\documentclass[sigplan]{acmart}
\usepackage{algorithm}
\usepackage[noend]{algpseudocode}
\algblockdefx{ParFor}{EndParFor}[1]{\textbf{parallel for} #1 \textbf{do}}{}
\algtext*{EndParFor}

\usepackage{listings}
\usepackage{booktabs}
\usepackage{microtype}
\AtBeginDocument{%
  }

\setcopyright{acmlicensed}
\copyrightyear{2026}
\acmYear{2026}

\setcopyright{none}
\renewcommand\footnotetextcopyrightpermission[1]{}

\acmConference[EGRAPHS '26]{}{June 15, 2026}{Boulder, CO, USA}

\begin{document}

\title{Rewrite System Showdown: Stochastic Search vs. EqSat}

\author{Qiantan Hong\textsuperscript{*}}
\email{qthong@cs.stanford.edu}
\affiliation{%
  \institution{Stanford University}
  \country{USA}
}

\author{Rupanshu Soi\textsuperscript{*}}
\email{rsoi@cs.stanford.edu}
\affiliation{%
  \institution{Stanford University}
  \country{USA}
}

\author{Yihong Zhang\textsuperscript{*}}
\email{yz489@cs.washington.edu}
\affiliation{%
  \institution{University of Washington}
  \country{USA}
}

\author{Alex Aiken}
\email{aaiken@stanford.edu}
\affiliation{%
  \institution{Stanford University}
  \country{USA}
}

\renewcommand{\shortauthors}{Hong, Soi, Zhang, and Aiken}

\newcommand{\TODO}[1]{\textcolor{red}{\textbf{TODO:} #1}}


\begin{abstract}
Equality saturation has become a dominant paradigm for equational program optimization.
However, it has never been rigorously compared to another approach to the same problem, even though several exist, the most notable being stochastic search.
In this paper, we compare equality saturation to stochastic search over five benchmarks to answer the question: are e-graphs actually good?
\end{abstract}

\maketitle

\renewcommand{\thefootnote}{\fnsymbol{footnote}}
\footnotetext[1]{Equal contribution.}
\renewcommand{\thefootnote}{\arabic{footnote}}

\renewcommand{\sectionautorefname}{Section}
\renewcommand{\subsectionautorefname}{Section}
\renewcommand{\subsubsectionautorefname}{Section}

\section{Introduction}

Equality saturation (EqSat) is an equational program optimization technique, with powerful, ready-to-use libraries such as egg \citep{egg}, egglog \citep{egglog,egglog-python}, and many others \cite{hegg,ego,scala-egg,metatheory}. It has seen success in many domains, from floating-point expressions \cite{herbie} to 3D printing \cite{szalinski}, from synthesizing better hardware \cite{esyn} to faster GPU kernels \cite{hardboiled}.
For many, equality saturation has become the go-to tool for building domain-specific program optimizers.

At a high-level, EqSat takes an input program and a set of semantics-preserving rewrite rules $R=\{l_1\rightarrow r_1,l_2\rightarrow r_2,\ldots \}$, searches the  space of equivalent programs defined by the rewrite rules, and returns the best program it finds. 
Importantly, EqSat uses the rewrite rules to grow an \emph{e-graph} of equivalent programs.
E-graphs compactly represent equivalence relations between programs, which allows EqSat to efficiently explore the search space.


Recently, a number of extensions to EqSat have been proposed, including DAG-costed extractions \cite{smoothe,tensat,glenn-treewidth}, effectful programs \citep{eggcc,numba-v2}, theories \cite{egraphmodulotheory}, optimistic analyses \cite{optimism}, and contextual reasoning \cite{assumenodes}.
The complexity of these extensions is often engendered by the compressed representation of terms in an e-graph.

Before we expend more effort in developing extensions, let us take a step back and ask: are e-graphs actually good? 
Somewhat surprisingly, EqSat has not been compared against other approaches to equational program optimization, not even in the original egg paper, but EqSat is certainly not the only approach to equational program optimization.
Because of the lack of a comparison, practitioners rely on popularity or word of mouth when picking an approach to program optimization.

To fill in this gap in the literature, we conduct the first extensive comparison of EqSat against another well-established technique: stochastic search.
Stochastic search, based on Markov chain Monte Carlo (MCMC) sampling \cite{mcmc-revolution}, has also found successful applications in a number of domains \cite{stoke,paramorphisms,floating-point,isa,taso}. Unlike equality saturation, stochastic search always manipulates concrete terms.

We implement stochastic search with an interface similar to the egg library, including two features that many EqSat applications rely on, namely, per-node analyses, and rules beyond purely syntactic rewrites.
This allows us to port existing EqSat applications to stochastic search with only mild effort.
We note that egg's interface is well-suited to host other equational program optimization techniques and is not limited to EqSat and stochastic search.

We collect a set of benchmarks for equational program optimization.
Our benchmarks range over different domains and from small, self-contained problems to full-blown applications taken from prior work.
The benchmarks are:

\begin{enumerate}
    \item Matrix chain multiplication
    \item Trigonometric simplification
    \item Indefinite integration
    \item Decompiling CAD into structured programs \cite{szalinski}
    \item Proving inequalities in the Halide compiler \cite{caviar}
\end{enumerate}

Preliminary results indicate that neither technique dominates the other, suggesting  that it may be possible to develop simpler alternatives to EqSat that are equally powerful for equational program optimization but avoid the complexity of e-graphs.

\section{Stochastic Search}
In this section, we give a brief overview of stochastic search.
Starting from an initial term $t$, stochastic search explores equivalent terms via Markov Chain Monte Carlo (MCMC) sampling.
The  search is governed by a proposal function $P(t)$, which induces a set of candidate terms through the application of the ruleset to any sub-term of $t$. 
A successor term $t' \in P(t)$ is sampled with a probability proportional to $\exp\left(-\frac{\beta}{2} (C(t') - C(t))\right)$
where $C$ is a cost function and $\beta \in \mathbb{R}$ is a hyperparameter called the inverse temperature.
The successor term replaces the current term and this process repeats until a computational budget is exhausted.

Restarting is crucial for the efficacy of stochastic search.
The distribution of solution quality across runs is often heavy-tailed, so many independent runs can give much faster search times than a few long runs \cite{koenig2021adaptive}.
Moreover, restarts help in getting a run out of local minima.
For these reasons, our implementation supports two kinds of restart.
In a soft restart, we set $\beta = 0$ periodically to help the search get out of local minima.
In a hard restart, we reset the state of the run to the very beginning; hard restarts only occur if the run has failed to make progress for a given number of steps.
Algorithm \ref{alg:stochastic-search} gives pseudocode for stochastic search, showing sampling and both kinds of restart.


\begin{algorithm}[t]
\caption{Stochastic Search}\label{alg:stochastic-search}
\begin{algorithmic}[1]
\Require Proposal function $P$, cost function $C$, initial term $t_0$, budget $B$, inverse temperature $\beta$, soft restart period $n_{\text{soft}}$, exploration steps $E$, hard restart period $n_{\text{hard}}$
\ParFor{$i = 1\ \textbf{to}\ B$}
    \State $n \gets 0$, $n_{\text{stall}} \gets 0$, $t \gets t_0$, $t^*_i \gets t_0$
    \While{$n_{\text{stall}} < n_{\text{hard}}$}
        \If{$n \bmod n_{\text{soft}} < E$}
            \State $\beta_i \gets 0$ \Comment{Exploration phase}
        \Else
            \State $\beta_i \gets \beta$ \Comment{Exploitation phase}
        \EndIf
        \State Sample $t'\in P(t)$ with probability $\propto$ \newline
           \hspace*{4em} $\exp\!\left(-\frac{\beta_i}{2}\left(C(t') - C(t)\right)\right)$
        \State $t \gets t'$, $n \gets n+1$
        \If{$C(t) < C(t^*_i)$}
            \State $t^*_i \gets t$, $n_{\text{stall}} \gets 0$
        \Else
            \State $n_{\text{stall}} \gets n_{\text{stall}} + 1$
        \EndIf
    \EndWhile
\EndParFor
\State \Return term with lowest cost from $\{t^*_1,\cdots,t^*_B\}$
\end{algorithmic}
\end{algorithm}


A central advantage of stochastic search is its embarrassingly parallel nature: we can spawn many threads, each with a different seed, to explore a different part of the underlying program space.
In fact, this scaling is critical to the performance of stochastic search (\autoref{sec:trig}).
Moreover, stochastic search only uses constant memory per thread as it only maintains a single term at a time.
This makes stochastic search scale well with time.
In contrast, not only is the parallelization of EqSat non-trivial and the subject of several research efforts \cite{parallel-eqsat, egglog}, the memory usage of EqSat can also blow up as more equalities are added to the e-graph, so any scaling of EqSat will always be constrained by available memory, unless mitigation measures like pulsing \cite{caviar}, guidance \cite{guidedeqsat}, or garbage collection are used.



\section{Showdown}
In this section, we present detailed results comparing EqSat and stochastic search for 5 applications.
Experiments are run on servers with a 64-Core CPU with hyperthreading and 500 GB of memory.
We give 10 seconds of wall-clock time to both techniques, except \autoref{sec:halide} where the time limit is set to 3 seconds.
We use egg's default \texttt{BackOffScheduler} for EqSat.
Unless noted otherwise, stochastic search is run using 128 threads, while EqSat is run single-threaded.
While this setup gives stochastic search 128$\times$ CPU time (but not wall-clock time), we believe it still constitutes a fair comparison due to the embarrassingly parallel nature of stochastic search.

\subsection{Matrix Chain Multiplication}
\label{sec:matmul}
This benchmark concerns the classic optimization problem of reassociating a chain of matrix products to reduce the total number of scalar multiplications.
To illustrate, consider a product of three matrices $ABC$, where the dimensions of the matrices are $2 \times 3$, $3 \times 4$ and $4 \times 5$ respectively.
There are two ways to associate this product, $(AB)C$ and $A(BC)$.
Recall that the number of scalar multiplications required to multiply two matrices of size $m \times n$ and $n \times k$ is $m n k$.
Using this formula, we can compute the number of multiplications needed in the two associations, which come out to be 64 and 90 respectively, making $A(BC)$ a much faster implementation of the chain multiplication.
This problem can be formulated as equational program optimization with two rules for the associativity of matrix multiplication.
We randomly generate tests with a number of matrices between 10 and 1,000.
For each test, we also compute the optimal association using the standard $\mathcal{O}(N^3)$ dynamic programming algorithm.

\begin{figure}[t]
    \centering
    \includegraphics[width=0.7\columnwidth]{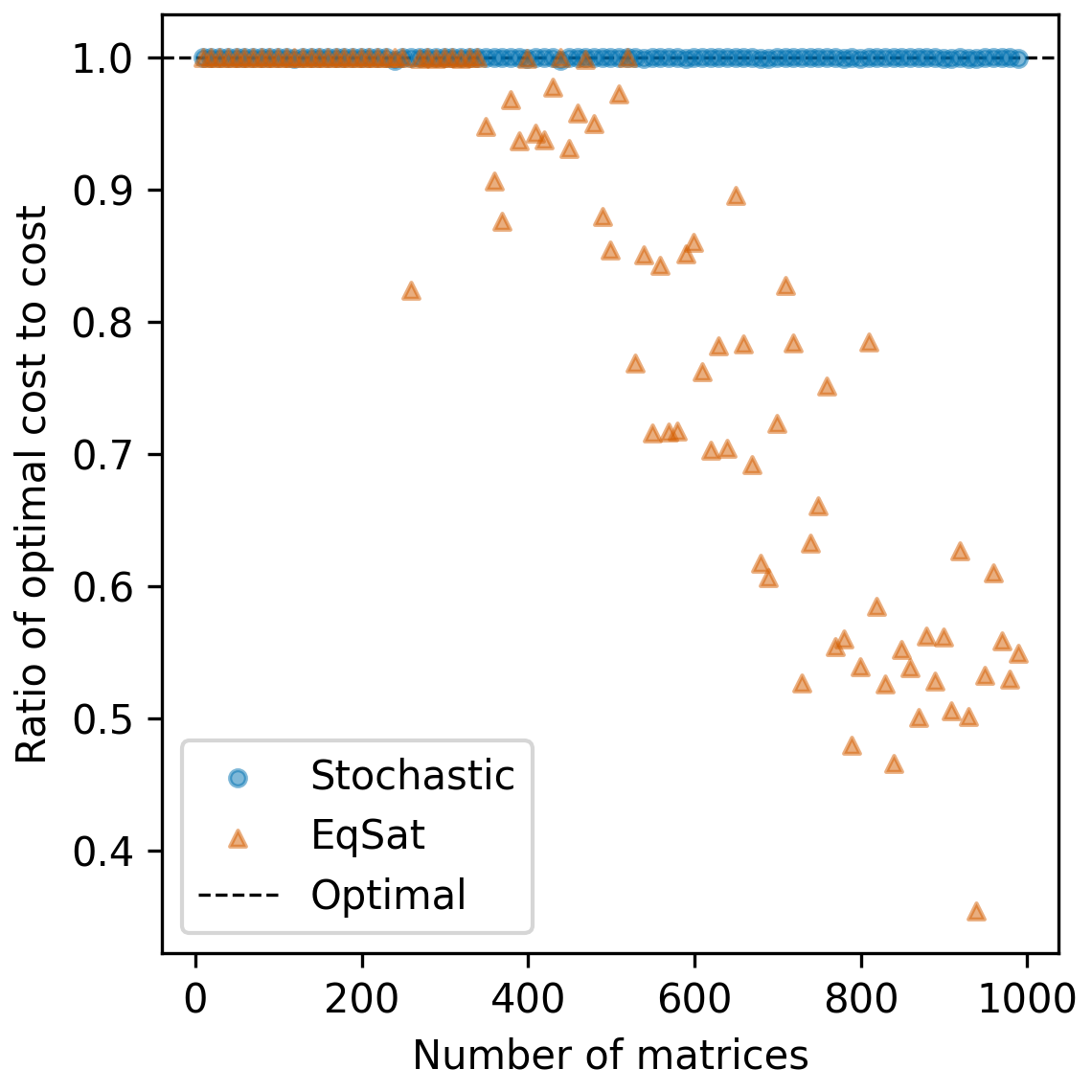}
    \caption{Showdown for matrix chain multiplication.}
    \label{fig:matmul}
\end{figure}

%
%
%

Results are given in Figure \ref{fig:matmul}.
The y-axis denotes the ratio of the optimal cost to the costs of associations produced by stochastic search and EqSat.
Stochastic search is always within 0.3\% of the optimal solution. 
EqSat is close to optimal when the number of matrices is moderate (e.g., $\leq 300$), but quickly becomes suboptimal as the number of matrices increases.

It is not surprising that, below a certain limit at least, both techniques fare equally well on this problem, despite the large program space.
It was recently proven that the mixing time of a random walk on the state space for stochastic search\footnote{The state space for a matrix chain of length $n$ is the 1-skeleton of associahedron $K_n$.} is $\mathcal{O}(n^3 \log^3 n)$, which is also the time for stochastic search to converge to its stationary distribution at high temperature \cite{assoc-mix}.
The space complexity of EqSat on this problem happens to be $\mathcal{O}(n^3)$---while not exactly the same, the two techniques are in the same ballpark regarding saturation or convergence.
However, as it takes EqSat $n$ iterations to saturate, saturation becomes intractable once $n$ gets larger, which explains the poor scalability.

We also tried pulsing for EqSat (omitted due to lack of space), where we run three iterations of EqSat, extract the best term from the e-graph, and use it to seed a fresh e-graph until the 10 s time limit is reached.
Pulsing significantly improves EqSat’s performance: for $n \leq 800$, it consistently remains within 1\% of the optimal cost.
However, as $n$ exceeds 800, performance begins to degrade again, with 5 out of 20 tests deviating by up to 10\% from optimal, and the worst case deviating by 65\%.


\subsection{Trigonometric Simplification}
\label{sec:trig}
For this benchmark, we collected a set of 35 trigonometric simplification exercises, such as proving that $\sin^4 x - \cos^4 x + 1 = 2 \sin^2x$.
The ruleset consists of trigonometric identities (e.g., $\sin^2 x + \cos^2 x = 1$) and purely algebraic ones (e.g., $x^2-y^2=(x+y)(x-y)$).
The cost of a term is the number of function symbols in the term, i.e., \verb|AstSize|.
There is an intended solution for each test, but it may not have the minimal cost.
We consider a test solved if a solution of cost less than or equal to the intended solution was found.

One difficulty in creating this benchmark is soundness.
In particular, two rules in our ruleset may cause unsoundness:
\begin{align}
\frac{b}{a} &\Rightarrow \frac{1}{\frac{a}{b}} \quad \text{if } b \neq 0 \tag{\textsc{recip}} \\
\frac{ab}{cb} &\Rightarrow \frac{a}{c} \quad \text{if } b \neq 0 \tag{\textsc{cancel}}
\end{align}

To understand why, we must reason about the domain of the LHS and the RHS of both rules.
By domain, we mean the set of values of the variables for which the LHS or RHS is defined,
e.g., $\operatorname{dom}(\operatorname{LHS}(\textsc{cancel})) = \{ (a, b, c) \, \vert \,  a, c \in T, b \in T \setminus 0 \}$ where $T$ is the set of all ground terms under consideration.
\textsc{recip} can cause unsoundness in both EqSat and stochastic search as
$\operatorname{dom}(\operatorname{RHS}(\textsc{recip})) \subsetneq \operatorname{dom}(\operatorname{LHS}(\textsc{recip}))$
and the guard only checks syntactic equality with zero.
Therefore, the rule will apply and result in unsoundness if, for instance, $b = x-x$ and 0 is not yet in the e-class of $x-x$ (for EqSat).
\textsc{cancel} can cause unsoundness in EqSat but not in stochastic search as
$\operatorname{dom}(\operatorname{LHS}(\textsc{cancel})) \subsetneq \operatorname{dom}(\operatorname{RHS}(\textsc{cancel}))$,
so it can cause unsoundness only if applied right to left, which happens in EqSat because EqSat forgets the direction in which a rule was applied.
Consequently, there is a proof in EqSat of $0 = 1$ by relying on \textsc{cancel} and a few other (sound) rules:
\[
1 \Leftarrow \frac{x-x}{x-x} \Rightarrow \frac{0}{x-x} \Rightarrow 0
\]
To ensure the validity of solutions in the presence of unsoundness, we adopt two mitigation strategies.
For EqSat, we follow a checkpointing strategy inspired by Herbie: a copy of the e-graph is made at the end of each iteration, and if unsoundness is detected, the run is aborted and the last checkpointed e-graph is used for extraction.
For stochastic search, the run undergoes a hard restart upon detection of unsoundness.
Across our benchmark suite, we ensure that solutions found by both techniques are always valid.

\begin{figure}[t]
    \centering
    \includegraphics[width=0.7\columnwidth]{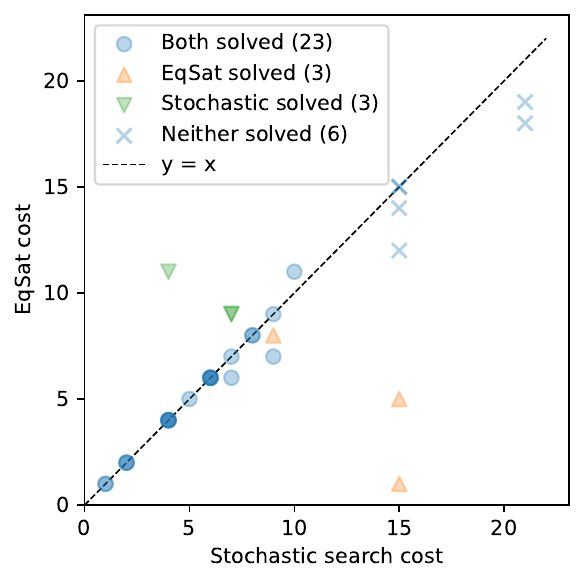}
    \caption{Showdown for trigonometric simplification.}
    \label{fig:trig}
\end{figure}

\begin{figure}
    \centering
    \includegraphics[width=0.7\linewidth]{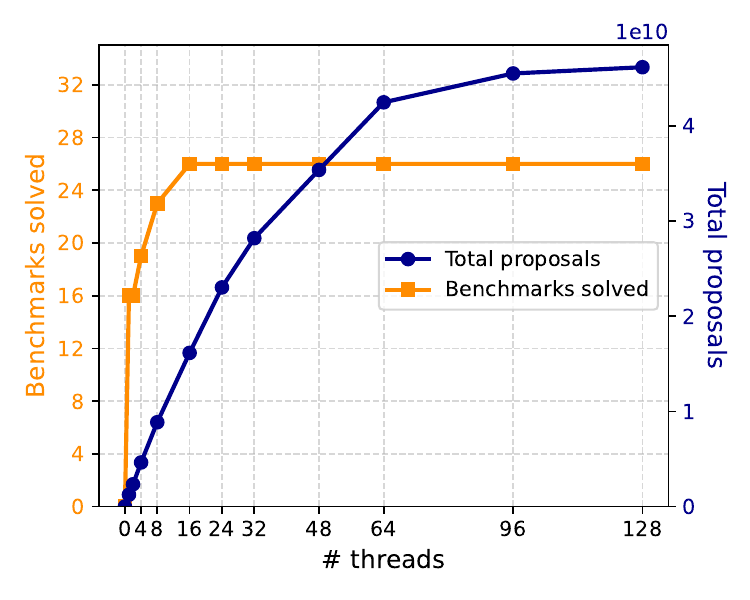}
    \caption{Number of proposals and benchmarks solved vs.~number of threads for stochastic search on trigonometric simplification.}
    \label{fig:scaling-proposal}
\end{figure}

Figure \ref{fig:trig} shows the results.
Both EqSat and stochastic search solved 26/35 tests, although the set of tests they solved are not the same.
We also measured how stochastic search scales with the number of threads, and the results are shown in \autoref{fig:scaling-proposal}.
We observe that the number of proposals (see Algorithm \autoref{alg:stochastic-search}) scales in proportion with the number of threads, and that stochastic search solves significantly more benchmarks with more threads, up to 16 threads.

\subsection{Indefinite Integration}

\label{sec:integ}

Also taken from homework exercises, this benchmark contains 10 problems of indefinite integration, such as proving that $\int x \cos x \, \mathrm{d} x = x \sin x + \cos x$.
The ruleset includes many rules found in the trigonometric benchmark, as well as rules specific to integration, such as for the integral of $\sin x$, or for integrating by parts.
Like trigonometric simplification, this ruleset can also cause unsoundness, and we deal with it by using the mitigation strategies listed in \autoref{sec:trig}.
Here we use different cost functions for EqSat and stochastic search.
For EqSat, we use \verb|AstSize| with the exception that the function symbols for integration and differentiation (needed for integration by parts) are weighed by 100 (other function symbols are weighed by 1).
However, this cost function is unsuitable for stochastic search as it makes applying linearity of integration too expensive: $\int (a + b) \, \mathrm{d} x \Rightarrow \int a \, \mathrm{d}x + \int b \, \mathrm{d}x$ would increase cost by 100, making this often fruitful step very unlikely to be taken.
To remedy this problem, we design a cost function that prefers integration of smaller terms.
The cost function for stochastic search weighs integration and differentiation nodes by the square of the costs of their children.
This function makes the linearity of integration a cost-decreasing step as $(x+y)^2 > x^2+y^2$.

\begin{table}[t]
    \centering
    \caption{Showdown for indefinite integration.}
    \label{tab:integ}
    \begin{tabular}{lc}
        \toprule
        Category & Count \\
        \midrule
        Both solved & 6 \\
        Only EqSat solved & 1 \\
        Only Stochastic solved & 2 \\
        Neither solved & 1 \\
        \bottomrule
    \end{tabular}
\end{table}


Another difficulty with constructing this benchmark was that many integration problems require a u-substitution, but neither technique supports u-substitutions, at least not without modifications.
In our benchmark, we did not include any problems that require a u-substitution.
To illustrate the problem, consider the integral $\int x^2 \sin x^3 \, \mathrm{d} x$.
The standard way to proceed is to substitute $u = x^3 \Rightarrow \mathrm{d}u = 3x^2 \mathrm{d} x$.
The integral becomes $\int \frac{1}{3} \sin u \, \mathrm{d} u = \frac{-1}{3} \cos u = \frac{-1}{3} \cos x^3$.
There are two difficulties with implementing this process.
First is the choice of sub-term to replace---this is a creative step and there are often numerous choices out of which only one or two will lead to the solution.
The wrong choice will further complicate the integral and it is difficult to recover once in this state, so an efficient method of undoing incorrect substitutions is desirable.
The situation is simpler for stochastic search, as threads that make the wrong substitution could simply restart, but for EqSat, it is unclear how incorrect substitutions could be undone efficiently.
The second difficulty is in the mechanical aspects of differentiating with respect to the chosen sub-term and backsubstituting into the integral.
In our example, the latter can be viewed as two rewrites that must happen at the same time: $x^3 \Rightarrow u, \, \mathrm{d}x \Rightarrow \frac{1}{3} \mathrm{d} u$.
Such a ``double rewrite'' is not supported by standard implementations of EqSat or stochastic search, though it can be added with some effort.

Results are shown in Table \ref{tab:integ}.
For this benchmark we do not give the costs as they are not particularly enlightening, due to the different weights of integration and differentiation function symbols.
Interestingly, of the 2 tests that stochastic search solved but EqSat did not, EqSat managed to carry out the integration but failed to simplify the resulting expression fully.
For the test that EqSat solved but stochastic search did not, stochastic search failed to carry out the integration fully.

\subsection{Decompiling CAD into Structured Programs}
\label{sec:szalinski}
\begin{figure*}
    \centering
    \includegraphics[width=1\linewidth]{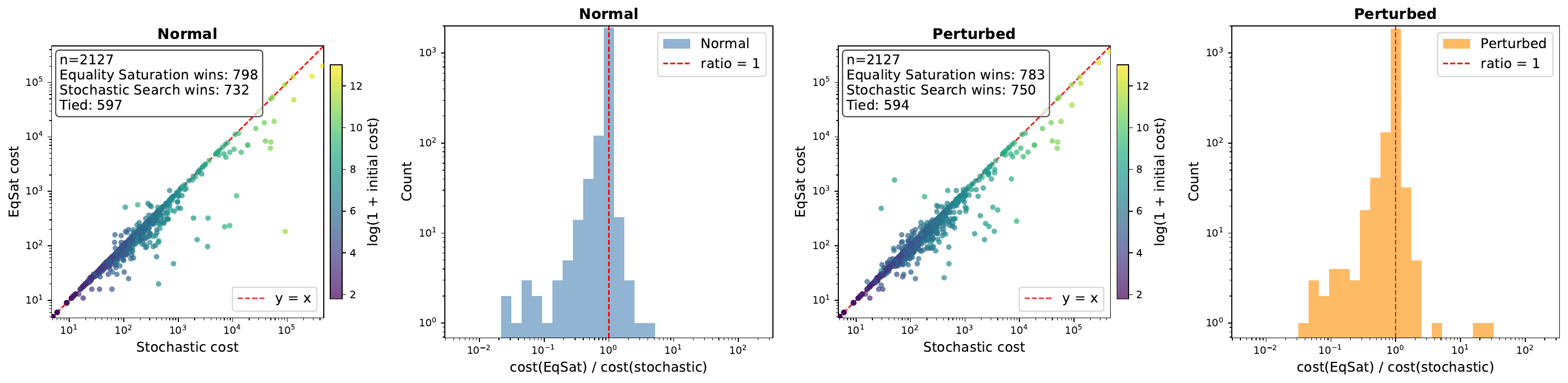}
    \caption{Showdown on compressing CAD expressions. }
    \label{fig:szalinski}
\end{figure*}


Next, we consider a decompilation task.
Szalinski \cite{szalinski} is a tool for compressing low-level unstructured CAD programs by finding equivalent programs that use functional primitives such as maps and folds.
Szalinski is especially attractive as a benchmark because it was also included as a benchmark in the egg paper.
The input to Szalinski is a flat CAD program like the following, which denotes 5 cubes placed in a straight line:


\begin{lstlisting}
(Union
  (Translate (0 0 0) Cube)
  (Translate (2 0 0) Cube)
  (Translate (4 0 0) Cube)
  (Translate (6 0 0) Cube)
  (Translate (8 0 0) Cube))
\end{lstlisting}

Using rewrite rules, Szalinski is able to identify repetitive structure in this program and synthesize a more compact equivalent, which is more amenable to user editing:
\begin{lstlisting}
(Fold Union
  (Tabulate (i 5)
    (Translate ((* 2 i) 0 0) Cube)))
\end{lstlisting}

We reimplement Szalinski using stochastic search and use the same set of rules as the original with small tweaks\footnote{We simplified the right-hand sides of two rules to make stochastic search's hill climbing easier. We added pushdown rules for \texttt{Sort} and \texttt{Unsort}. Finally, we added rules for rewriting the intersection/union of a singleton into itself.}.
We evaluate the two implementations on the Thingiverse dataset.
As in the Szalinski paper, we consider both the original programs from Thingiverse and their perturbed variants; the latter are meant to test robustness of the decompilation.

Results are shown in \autoref{fig:szalinski}.
The left two figures show the evaluation results on the normal programs, and and right two figures show results on perturbed programs.
The scatter plots use a similar format as before: the x-coordinate of a point gives the cost of the program found by stochastic search, and the y-coordinate gives the cost found by EqSat.
The histograms show the distribution of the ratios between costs found by EqSat and by stochastic search.

Overall, neither technique dominates the other---there are programs on which EqSat finds a better solution than stochastic search, and programs on which the opposite is true.
However, we do find that on programs on which stochastic search wins, it often wins by a small margin, whereas on programs on which EqSat wins, the margin is usually larger.
We suspect this is because some of Szalinski's rewrite rules are ill-suited to stochastic search, which causes it to miss certain critical rewrite paths..
To illustrate Szalinski's workload, consider an abstract rewrite system with three rules: $\{a \Rightarrow  b, b\Rightarrow a,  \, f(b, \ldots, b) \Rightarrow g(b, \ldots, b)\}$, where $f$ and $g$  have some large arity $N$.
We start with the term $f(a, \ldots, a)$ and our goal is to prove it equal to $g(b, \ldots, b)$.
Further assume that the cost function is simply $C\left(g(b, \ldots, b)\right) = 0$ and 1 otherwise.
EqSat can prove this within 2 iterations.
On the other hand, because the cost function provides little guidance, stochastic search must bounce around this high-dimensional space until it hits the term $f(b, \ldots, b)$ purely by chance, which requires, in expectation, $\mathcal{O}(2^N)$ time.
Only then can the rule $f(b, \ldots, b) \Rightarrow g(b, \ldots, b)$ apply, opening up a path to the solution.
Though a simplified example, essentially the same thing happens in Szalinski, as all the \verb|Translate| function symbols need to be in a certain form before they can be folded together.
It is possible that the performance of stochastic search on this benchmark can be improved with rule engineering and a better cost function, but we leave that as future work.



\subsection{Proving Inequalities in the Halide Compiler}
\label{sec:halide}

\begin{table}[t]
    \centering
    \caption{Showdown for proving inequalities from Halide.}
    \label{tab:halide}
    \begin{tabular}{lc}
        \toprule
        Category & Count \\
        \midrule
        Both solved & 2298 \\
        Only EqSat solved & 361 \\
        Only Stochastic solved & 184 \\
        Neither solved & 2943 \\
        \bottomrule
    \end{tabular}
\end{table}

\begin{figure}
    \centering
    \includegraphics[width=0.7\linewidth]{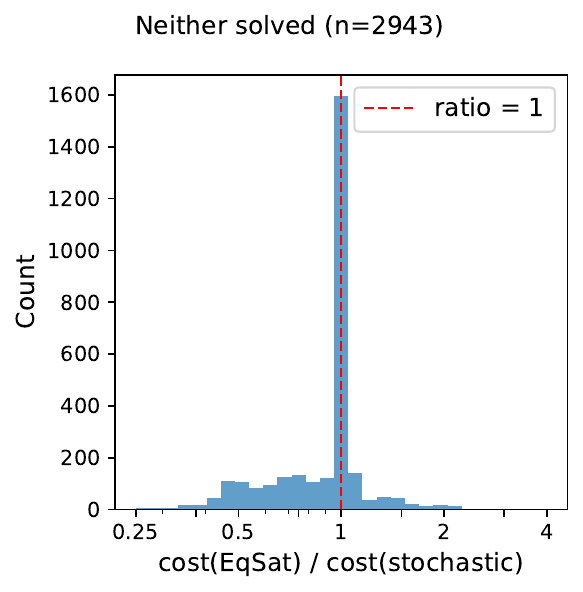}
    \caption{Cost distribution of unproved Halide inequalities.}
    \label{fig:caviar}
\end{figure}

The last benchmark is from Caviar \cite{caviar}, an EqSat reimplementation of the rewriting system inside the Halide compiler.
Halide generates many inequalities during its bounds inference, and proving them is critical for Halide to generate efficient code.
These inequalities concern algebraic expressions, similar to our benchmarks on trigonometry and integration, as well as max and min operators.
We implemented the same set of rules with stochastic search, and evaluate both systems on Caviar's ``hard'' dataset.
Each test consists of an inequality over symbolic variables to be proved or disproved, e.g., $\max(i, 2) < \max (i+3, 3)$.

Results are shown in \autoref{tab:halide} and  \autoref{fig:caviar}.
Of the 5,786 tests, 40\% were solved by both techniques, 6\% were solved by EqSat only, 3\% were solved by stochastic search only, and a large fraction, about 50\%, were not solved by either.
Note that Caviar's ruleset is relatively weak, and therefore, not all tests can be decided using it.
Overall, the results on Caviar align with the other benchmarks: while there is a core set of tests passed by both techniques, the tests passed by EqSat are neither a superset nor a subset of the tests passed by stochastic search, and vice versa.

\section{Discussion}
Finally, we discuss the differences between EqSat and stochastic search in light of our results, draw some conclusions, offer some ideas for future work, and answer the question raised in the introduction.

\subsubsection*{E-graphs good?}
Anecdotally, while working on this project, we were often surprised by how well EqSat performed on some benchmarks.
For matrix chain multiplication in particular (\autoref{sec:matmul}), we expected stochastic search to comfortably outperform EqSat for medium to large problems, as a long sequence of rewrites may be required to reassociate the chain into the optimal form and finding this chain ought to be difficult without any cost guidance.
However, the results indicate that EqSat performs almost optimally for problem sizes up to 300, which is already large.
This can be attributed to the compression provided by the e-graph---although the size of the search space is exponential in the number of matrices, it can be represented by the e-graph in just $\mathcal{O}(n^3)$ space.

While we still believe that EqSat will perform poorly for problems that require long sequences of rewrites, and some evidence to the affirmative is provided in \autoref{sec:matmul}, we had a difficult time coming up with a realistic example of such a problem.
It may be the case that EqSat works well for exactly the kinds of equational program optimization problems that occur in practice.

\subsubsection*{Is EqSat Pareto optimal over stochastic search?}
A common theme across our 5 benchmarks is that neither technique comes out to be Pareto optimal over the other---there are tests on which EqSat performs better than stochastic search, and vice-versa.
This result suggests that a combination of the two techniques, perhaps EqSat imbued with some stochasticity, or a variant of stochastic search that uses an e-graph to do ``local'' exploration while relying on randomness to jump to far parts of the space, could be the ultimate equational program optimization engine.

\subsubsection*{Unsoundness and directionality of rules.}
One difference between the two techniques is that a directed rule induces an equivalence relation over terms in the case of EqSat and a refinement relation over terms in the case of stochastic search.
As an upshot, some rules that cause unsoundness for EqSat work just fine for stochastic search, such as $\frac{x}{x} \Rightarrow 1$.
\autoref{sec:trig} discusses how we mitigate this issue for some benchmarks.
Anecdotally, unsoundness was a larger issue for EqSat than for stochastic search---the latter had to be run for much longer before it exploited unsoundness.
It is unclear to us whether this is an advantage or disadvantage of EqSat, because one could argue that quick exploitation of unsoundness indicates better ability at exploiting the ruleset.

\subsubsection*{Hyperparameters.}
A disadvantage of stochastic search is the requirement of hyperparameter tuning, namely, of tuning the restart schedule and values of $\beta$.
While it was not difficult to find good values of these hyperparameters in our experiments, no such tuning had to be done for EqSat.
It is however worth noting that several applications of EqSat do require hyperparameter tuning in the form of rule scheduling.
Rule scheduling is used when applying the entire ruleset at the same time would lead to a blow-up of the e-graph, so the ruleset is divided into disjoint sets and applied to the e-graph in phases.
\verb|BackoffScheduler| in egg is a simple example of rule scheduling.
Another hyperparameter in EqSat is the concept of top-k e-nodes, which is used in incremental EqSat \cite{incremental} to handle very large e-graphs.

\subsubsection*{Scalability.}
An advantage of stochastic search is scalability owing to its embarrassingly parallel nature.
\autoref{fig:scaling-proposal} provides some evidence to this effect by showing how the number of passing tests scales with the number of threads given to stochastic search.
In principle, we could scale stochastic search up to thousands or even tens of thousands of threads by using multi-node execution, whereas parallel or distributed execution at this scale remains challenging for EqSat. 

\subsubsection*{Non semantics-preserving rewriting.}
An advantage of stochastic search that we did not explore in this paper is that, unlike EqSat, it can benefit from non-semantics-preserving rules.
Perhaps the most well-known application of this is \textsc{stoke} \cite{stoke}, a superoptimizer based on stochastic search for x86 programs.
Most of the rewrites in \textsc{stoke} were not semantics-preserving; instead, its cost function had a correctness component that guided its search to correct programs.
\textsc{stoke} often found completely different assembly-level implementations of a given algorithm, and this was possible only because its rewrite rules allowed it to travel to very different parts of the space from which it started.
It would be very difficult to match these results using only semantics-preserving rewrites.


\begin{acks}
Benjamin Driscoll, Chris Gyurgyik, AJ Root, and Rohan Yadav provided helpful feedback on this manuscript.
\end{acks}

\bibliographystyle{ACM-Reference-Format}
\bibliography{references}

\end{document}